# Deep Learning-Assisted Simultaneous Targets Sensing and Super-Resolution Imaging

*Jin Zhao[1], Huang Zhao Zhang[2], Ming-Zhe Chong[1], Yue-Yi Zhang[1], Zi-Wen Zhang[1], Zong-Kun Zhang[1], Chao-Hai Du[1], and Pu-Kun Liu[1], **

*[1]State Key Laboratory of Advanced Optical Communication Systems and Networks, School of Electronics, Peking University, Beijing, 100871, China*

*[2]School of Computer Science, Peking University, Beijing, 100871, China*

*Correspondence to Pu-Kun Liu: pkliu@pku.edu.cn

**ABSTRACT:** Recently, metasurfaces have experienced revolutionary growth in the sensing and super-resolution imaging field, due to their enabling of sub-wavelength manipulation of electromagnetic waves. However, the addition of metasurfaces multiplies the complexity of retrieving target information from the detected fields. Besides, although the deep learning method affords a compelling platform for a series of electromagnetic problems, many studies mainly concentrate on resolving one single function and limit the research's versatility. In this study, a multifunctional deep neural network is demonstrated to reconstruct target information in a metasurface-targets interactive system. Firstly, the interactive scenario is confirmed to tolerate the system noises in a primary verification experiment. Then, fed with the electric field distributions, the multitask deep neural network can not only sense the quantity and permittivity of targets but also generate super-resolution images with high precision. The deep learning method provides another way to recover targets' diverse information in metasurface-based target detection, accelerating the progression of target reconstruction areas. This methodology may also hold promise for inverse reconstruction or forward prediction problems in other electromagnetic scenarios.

## 1. INTRODUCTION

Relying on the principles of scattering, diffraction, and other interactions, sensing and imaging techniques in the microwave, terahertz, and optical regions accommodate crucial tools for indicating the information of the measured target. Sensing is widely deployed in a large number of applications,[1] such as the measurements of refractive indices,[2] distance or thickness,[3-4] movement of human bodies,[5-6] the composition of materials,[7] and so on. On the other hand, imaging technologies encompass medical or biological tissue inspection,[8-9] integrated circuits flaw detection,[10] and so forth. In brief, sensing and imaging processes usually perform in a measurement system where electromagnetic fields interact with the targets.[11]

Metasurfaces are constructed surfaces that are composed of subwavelength atoms periodically or non-periodically.[12] Due to their inherent abilities to manipulate electromagnetic waves in subwavelength footsteps, metasurfaces have attracted continual attention in the sensing and imaging measurement systems.[11, 13-15] For sensing, the metasurfaces could support high-Q Fano resonances, allowing to monitor shifts of the resonance frequencies when the metasurfaces are bonded with the thin analyte.[16] Accordingly, the refractive index properties of the analyte can be demonstrated by the movement of the resonance peak.[16] Besides, with the assistance of metasurfaces, the imaging resolution manages to be enhanced or even surpass the confinement of the classical diffraction limit (approximately half of the wavelength[17]). One general strategy to circumvent the diffraction limit is employing surface waves to irradiate the object. Then the fine information carried by a shifted spectrum can propagate to the detection plane, implementing subwavelength-resolution imaging.[18] Structural illumination microscopy (SIM) is a typical application of the spatial frequency spectrum shift method.[19-20] In recent SIM studies, metasurfaces[21-22] are prone to guide the surface waves and provide the required illumination.

Although metasurfaces are prevalently applied in diverse measurement systems and promote flexibility and freedom, they further increase the solution complexity. Taking SIM imaging as an example, the post-processing algorithms tend to produce confusing imaging artifacts with the influence of extra noise yielded by metasurfaces. Furthermore, the introduction of metasurfaces is usually accompanied by recovering higher-resolution images at the cost of more raw images,[23] which may lead to laborious post-process procedures. Besides, the SIM method applying metasurface requires a single-harmonic pattern so the iterative Gerchberg-Saxton (GS) algorithm is suitable.[24] This method is expired with the metasurface-supported muti-harmonic pattern which can bring more fine information once a time. Thus, there is an imminent demand for a high-precision and concise approach to deal with the metasurfaces-assisted system to reconstruct the target information.

Machine learning, especially deep learning (DL), is a potent nonlinear algorithm typically stacking and linking diverse neural network (NN) architectures. Numerous activation cells are present in each layer of NN, and they are computed using the cells from the preceding layer and the connection weights between the two layers (in addition to the first data-input layer)[25]. Driven by data, the connection weights and other network hyperparameters can be adjusted continuously to accomplish the ultimate goal. To be specific, for supervised DL, this optimization (or training) process is implemented until the loss function (the difference between the network's output and the ground truths) is reduced to a convergence value.[25] When the network is fixed, it can be tested on the remaining dataset to predict the required data. Due to its dramatic ability in data treating, DL has recently been investigated by researchers for a surge of applications including metasurfaces unit inverse design,[26-28] spectral prediction,[29-30] computational imaging,[25] and objects detection[3]. Nevertheless, most of the well-behaved DL architectures addressing electromagnetic problems mainly concentrate on a single goal, such as recovering high-resolution pictures from low-resolution images,[31] obtaining frequencies spectrum according to the device geometry or vice versa,[26-30] or recognizing (classifying) different kinds of input images.[32-33] Although DL yields excellent results in the aforementioned studies, further exploiting its potential versatility is of great significance.

In this work, a multitask DL network is proposed to synchronously acquire the quantity, permittivity, and super-resolution images of the targets in a metasurfaces-involved electromagnetic system. When the metasurface is exposed to a plane wave illumination, both the surface harmonic waves on the metasurface and the partially permeable plane wave interact with the targets and then scatter to the total field. Since several wave components overlap in our illumination pattern, the general formula derivation or the iterative restoration algorithms for SIM may be cumbersome and ill-suited. Thus, a multitask DL network is presented to reconstruct multiple targets' information, with only one intensity frame as the input data. After repeatedly 700 training epochs, the final model realizes the prediction of targets' quantity and permittivity with an accuracy of approximately 100% and 95%, respectively. The output images reach high precision with more than 42dB peak signal-to-noise ratio and an approximate $0.2\lambda$ resolution. To some extent, this method decreases the complexity introduced by the metasurface, including the artifacts and multi-frame inputs. Besides, this method is compatible with complex multi-harmonic illumination patterns, which is essentially realized with the nonlinear mapping idea. Aside from the super-resolution reconstruction, quantity and permittivity can also be sensed with the multitask DL network, verifying the multifunction of this method. At last, a new DL network is further demonstrated to forward predict the electric field distribution. In summary, constructing a DL network can provide a promising solution to some cumbersome electromagnetic problems.

## 2. METHODS

In this task, the metasurface with one-dimension (1D) periodic circular metal grating is adopted to provide multiple illumination electromagnetic waves. The multiple illuminations interacting with targets are composed of two opposite surface waves and one propagation wave component. Subsequently, we capture the electric intensity field distribution in a line away from the targets and equip it as the input data of the DL model. There are three tasks in our DL model, whose labels (or ground truths) are quantity, permittivity, and super-resolution image of the targets. The DL model is fed with the intensity patterns and learns to approach the corresponding ground truths for each task. Finally, all the aforementioned information can be predicted by the DL model. The schematic of this presented method is exhibited in Figure 1.

**2.1. Metasurface-based illumination system.** In the proposed approach, a circular metal grating displayed in Figure 2a is employed as the unit structure. The dimensions of the element are $P = 16.7$ mm, $R_1 = 6.67$ mm, $R_2 = 5.17$ mm, $t = 0.508$ mm, and $\theta = 15^\circ$. When the unit is aligned in the $x$-direction with period $P$, the vertical incident plane wave can be modulated and diffracted by the array. A further understanding of this periodic structure is according to the Floquet theorem.[34-36] The incident wave has a zero-value wavenumber component along the $x$-direction, while the periodic system can shift the incident tangential ($x$-direction) wavenumbers with $2n\pi/P$ ($n = \pm1, \pm2, \pm3\ldots$). In this condition, the final waves sustained in the metasurface are a series of spatial harmonics, each of which has different tangential wavenumbers manifesting as $\beta = \beta_0 \pm 2n\pi/P$ ($\beta_0 = 0$, $n = 1, 2, 3\ldots$). Simulating the metasurface system in commercial software COMSOL

Multiphysics, the $x$-direction electric fields ($E_x$) are obtained as illustrated in Figure 2b. The complex field detection line is intensely close to the metasurface (the white dotted line in Figure 2c) so that the surface wave components can be included. It is noted that the corresponding spatial frequency spectrums are also acquired through the spatial Fourier transform. As illustrated in Figure 2d, there mainly exist three components with the tangential wavenumbers being 0 and $\pm 2\pi/P$, verifying the verifying the periodic modulation feature of the surface array. Furthermore, Figure 2e,f demonstrates the dominating tangential wavenumbers are still 0 and $\pm 2\pi/P$ when the period $P$ is altered.

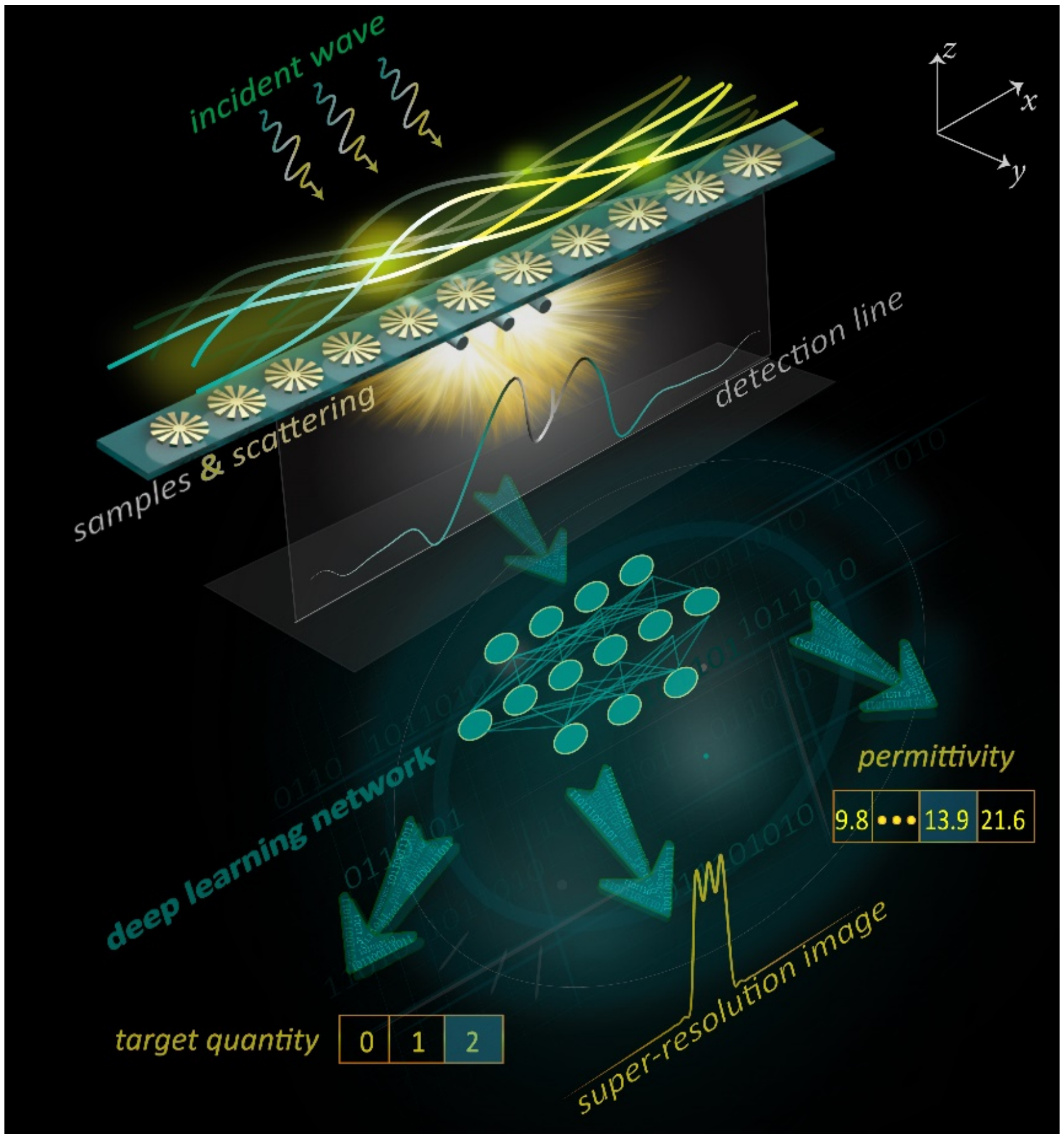

**Figure 1.** Schematic of the deep learning-enabled target reconstructions based on the metasurface-modulated wave sources.

After clarifying metasurface-supported modes under the vertical incident plane wave, the targets are placed and almost attached to the metasurface. In Figure 2g,h, ceramics cylinders are utilized as the targets and interact with the electromagnetic waves on the metasurface. The scattered fields will overlap with each other, establishing the final total field in the intensity detection line. The proposed intensity-based target reconstruction is obstructed by the multi-harmonic illumination and interaction patterns. Attempting to discover the implication relationship between the intensity along the detection line and the targets' information, we yield the DL network to meet this challenge.

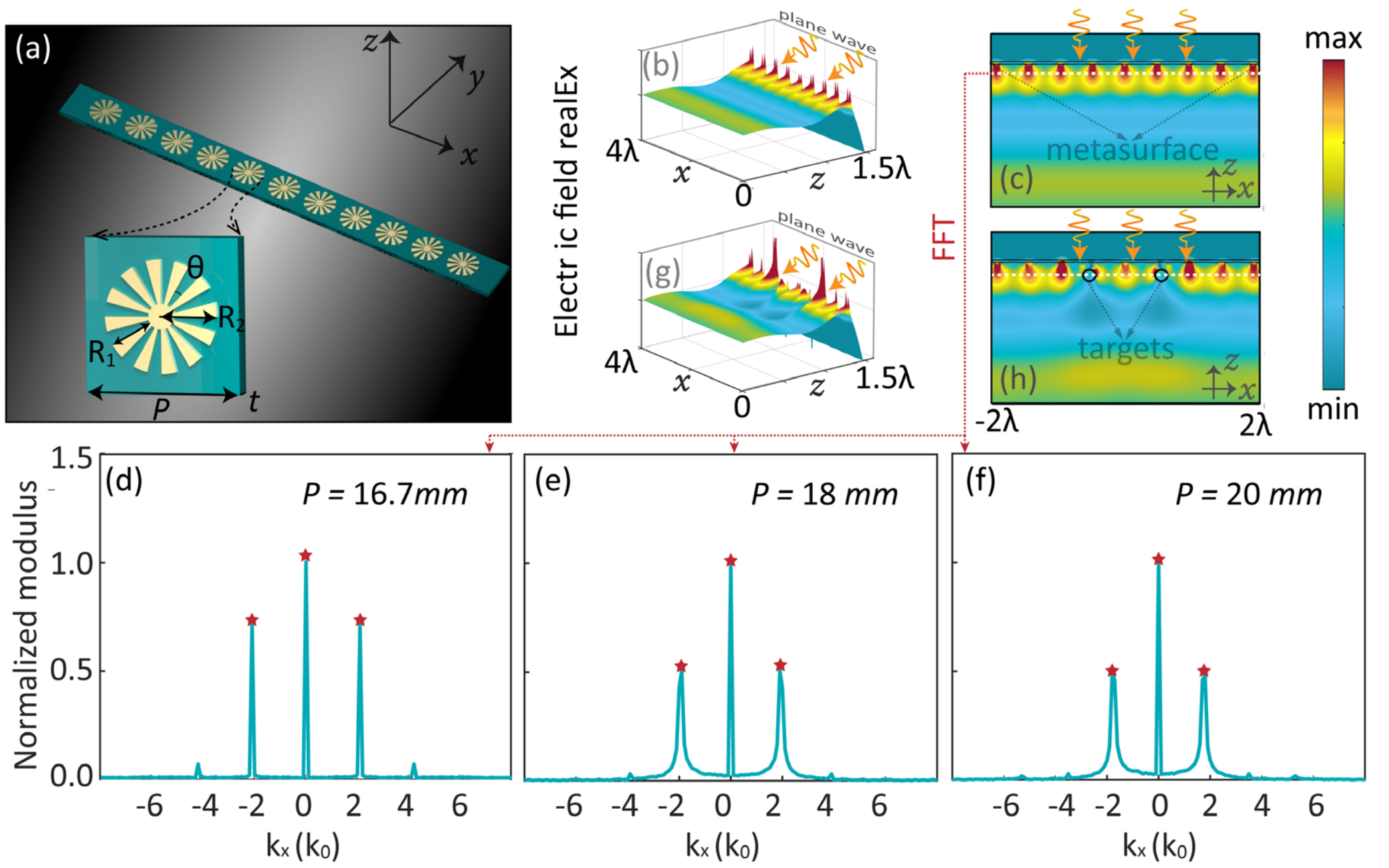

**Figure 2.** Schematic of the metasurface structure. (a) The periodical structure is composed of a metal circular grating unit, where $P$ = 16.7 mm, $R_1$ = 6.67 mm, $R_2$ = 5.17 mm, $t$ = 0.508 mm, and $\theta$ = 15°. (b), (c) Respective three-dimensional (3D) and two-dimensional (2D) electric field distribution of the real $x$-component. (d)-(f) Spatial Fourier transform of $E_x$ distribution with different unit period $P$. The electronic field $E_x$ is detected along the white line position in (c). There is only the metasurface and no targets when detecting the electronic field. (g), (h) 3D and 2D real $E_x$ distribution when two targets are located after the metasurface.

**2.2. Data acquirement.** The constructed DL model belongs to supervised learning, which requests both the input data and the supervisory signals. The input data is captured in the simulation system, that is the intensity $E_x$ field on the detection line under different targets. The monitoring signals contain the actual quantity, permittivity, and super-resolution images of the targets.

With the assistance of the commercial software COMSOL Multiphysics, a total of 5100 sets of data are acquired. Each pair of data contains the intensity field and the corresponding targets' information. Single, double, and triple cylindrical targets are simulated with their relative permittivity varying from 10 to 20. For the single target, its position ranges from $-0.49\lambda$ to $0.49\lambda$. For the case of two targets, their central distance increases from $0.2\lambda$ to $2\lambda$, which includes the super-resolution status. Besides, the shortest distances between any two points are also $0.2\lambda$ in the three targets distribution.

For each target's arrangement, we observe the intensity fields of $E_x$ along the line about $1.3\lambda$ away from the targets. The length of the detection line along the $x$-direction is $10\lambda$ and the sampling interval is $0.5\lambda$, that is, per detection can obtain 21 discrete values. Seeking to contribute abundant information for the DL model, we fit the original 21 points curve into 701 points curve as the eventual input data. This method is of significant advantage in speedy and straightforward input data acquisition since one frame (an intensity curve) is the only input for a single reconstruction. In a contrast, excellent works about SIM usually demand multiple raw images under vertical, oblique, and surface wave illuminations for one single retrieval[37]. Resemble situation is that the working frequencies are constantly altering to obtain a series of information for one rebuilding[24]. Thus, the one-frame detection for target restoration in this work may be worth promoting.

The quantity, permittivity, and super-resolution image of the targets are the supervisory signals for the corresponding input intensity fields. The quantity and permittivity of targets can directly refer to the settings in the simulation. While by the superposition of the Airy disks, the super-resolution images are obtained. When there are multiple targets, several Airy disks are superimposed to approximate the corresponding super-resolution images. It is worth noticing that the supervisory signal may also be estimated by other similar functions such as the Gaussian function. Besides, the FWHM and the minimum distance of two peaks depend on the expected image resolution.

The 5100 simulated datasets are randomly divided into training, validating, and test datasets, with the proportions being 80%, 10%, and 10%, respectively. The connection weights of the DL model are trained by the training dataset and the validating data is exploited to monitor the model. In the end, the model's performance is tested on the test dataset.

**2.3. Experimental validation.** To increase the convincingness of the system, we experimentally verify the intensity field distribution in this irradiation scenario. It should be reminded that the purpose of the experiment is to demonstrate whether or not the targets (cylinders) can break through the background noise in experiments and affect the field distribution. In simulations, when a cylindrical sample is attached to the system, the intensity distribution in the detected line is affected by the sample, so diverse intensity curves are obtained. The variety of intensity curves owing to the introduction of cylindrical samples is an exceedingly critical step in the next DL procedure. To a certain extent, the DL model reconstructs the corresponding information of the targets by learning the fluctuation of the intensity curves. Thus, if the sample is unable to significantly disturb the intensity curve, the ensuing DL network's performance will suffer from insufficient information. However, in the real experimental scene, the noise in the system (such as the reflections from the transmission line, the detection probe, and the interference from the experimental fixtures) may cover the field disruption caused by the cylinder. In this instance, the cylindrical sample's impact on the field distribution cannot be distinguished. The difficulty of subsequent DL recovery will significantly rise once this phenomenon occurs. To confirm the impact of the cylinder on the intensity field, we set up the experimental schematic system as depicted in Figure 3a. To approximately imitate the incident plane wave, a horn antenna is applied in the experimental apparatus to produce a Gaussian beam. The beam is impacted on the metasurface vertically. We deploy a probe 30–50 mm beyond the metasurface to measure the distribution of the $x$-direction electric field. The probe's detection range is roughly 24 cm (almost $7\lambda_0$ at 8.6GHz), and the detection step is around 4 mm. A perforated 3D-printed resin plate is employed to support the cylindrical sample.

Figure 3b,c illustrates different views of the experimental setup, where the metasurface and the sample are fixed in the supporting platform. A probe is scanned to detect the field distribution along the dot black line. The measured intensity of the $E_x$ field is plotted in Figure 3d. It can be seen that when the position of the sample is randomly moved three times, the peak distribution of the curve shifts accordingly. This phenomenon indicates that the system noise does not overwhelm the influence of the samples in this experimental situation.

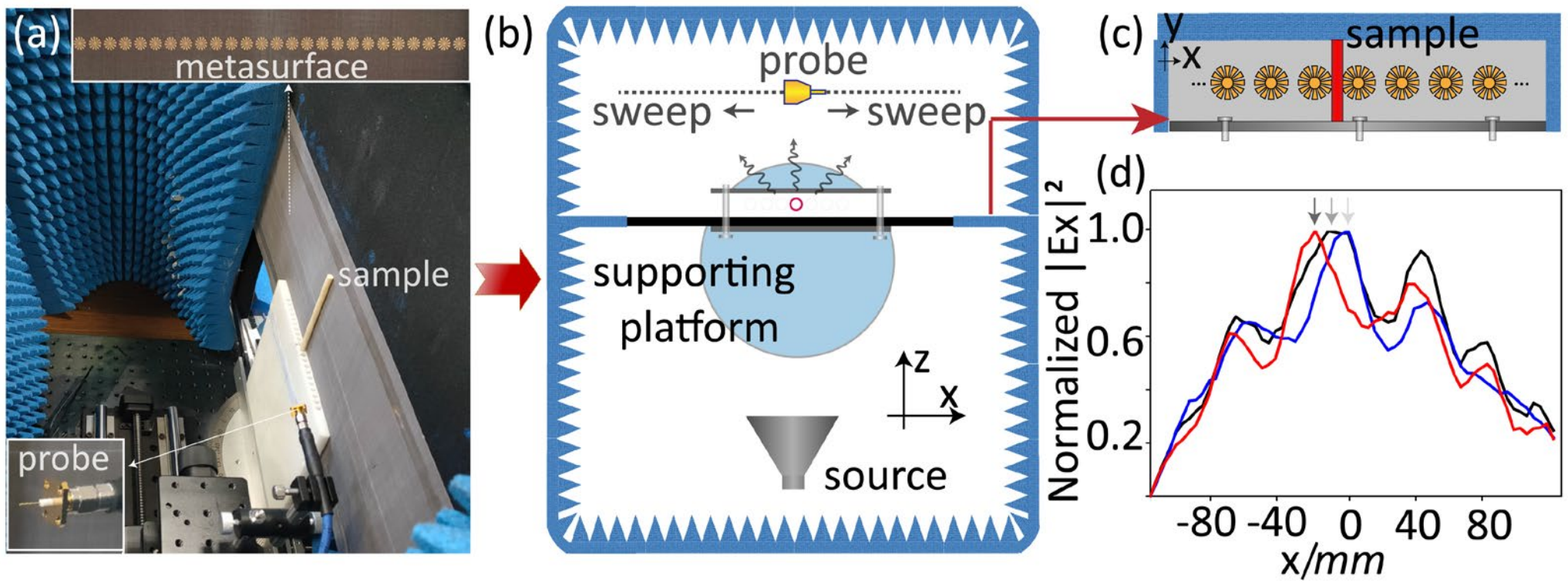

**Figure 3.** Experimental setup of the verification. (a) Picture of the experimental configuration, in which the ceramic sample is erected behind the metasurface. (b) Top view of the experimental setup. A horn antenna is applied as the wave source, and the metasurface is fixed in front of the sample by the supporting platform. The probe scans along the sweep line to detect the *x*-direction electric field distribution. (c) Front view of the metasurface. (d) The detected electric field with different sample positions.

**2.4. Deep learning neural network.** The DL model consists of the encoder, decoder, classifier, and regressor as demonstrated in Figure 4. Initially, the encoder is fed with the input intensity data. Following the triple-layer convolution processing in the encoder, the output becomes a tensor of size 256*3. This tensor is supplied to the subsequent decoder, classifier, and regressor, respectively. The decoder performs three cascaded trans-convolution calculations on this 256*3 tensor and thereafter produces a vector to approximate the real super-resolution image. As for the classifier, it is a multilayer perceptron (MLP) encompassing four fully connected (FC) layers. Operating on the input tensor, the classifier will output a probability distribution of the targets' quantity, that is, the probabilities of the targets' number being 1, 2, and 3. Logically, the predicted probability distribution is required to approach the truth (or labels in the following description). The regressor is also an MLP with four FC layers, whose output is a real number as the approximate estimate of the targets' permittivity. In brief, the encoder is shared by the three subsequent modules with different reconstruction tasks. Owing to this branched network architecture, the model can simultaneously output target sensing and super-resolution imaging information.

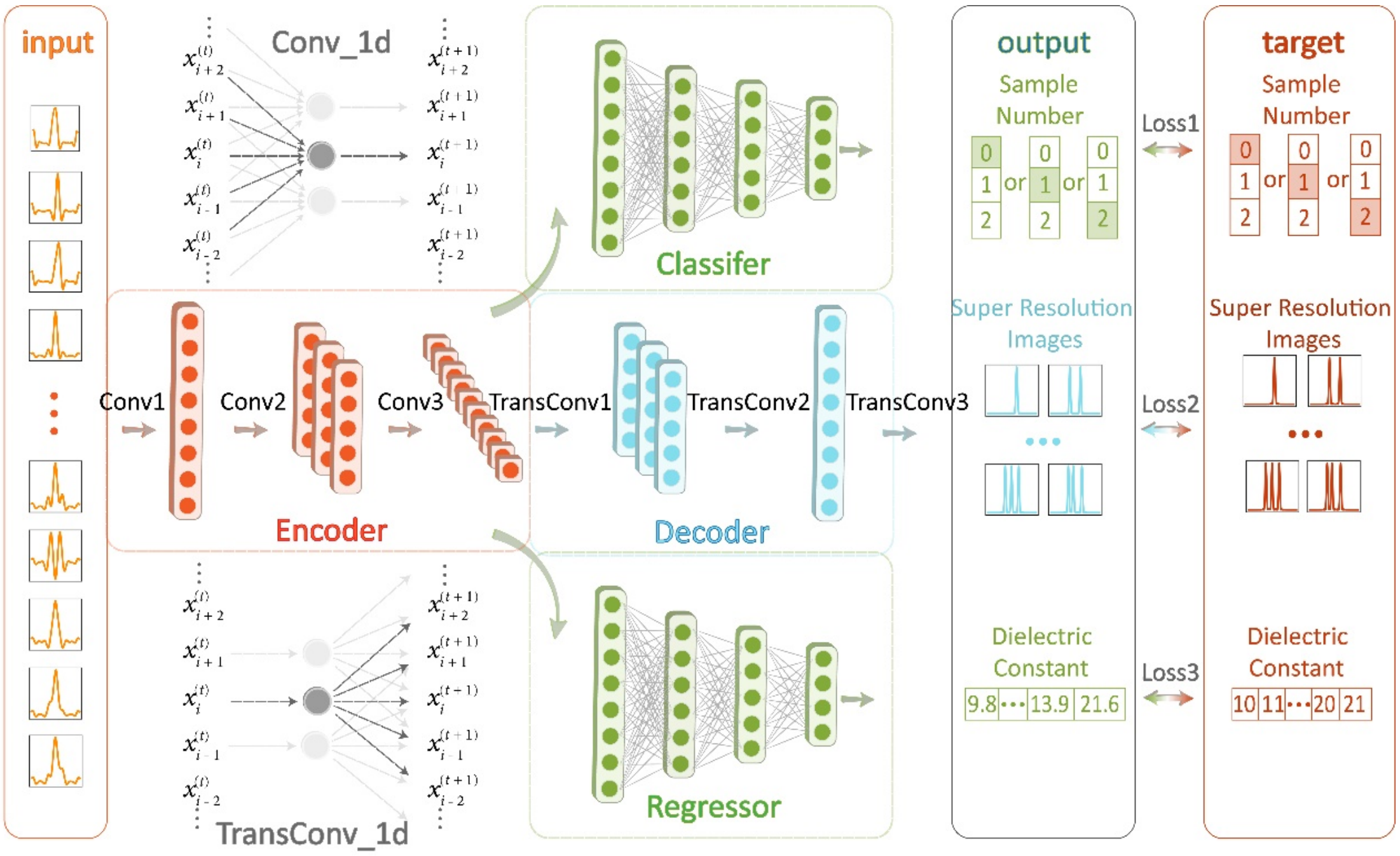

**Figure 4.** The proposed deep learning network structure for the simultaneous prediction of the targets' quantity, permittivity, and super-resolution image.

Next, the difference between the model's output and the actual supervisory signal is evaluated by the loss function. By minimizing the loss function during the model training process, it is promising to attain high precision and accuracy in prediction performance. In this work, the peak signal-to-noise ratio (*PSNR*) is calculated to monitor the difference between the model-predicted super-resolution images and the practical super-resolution images. *PSNR* (dB) is computed as following:[38]

$$PSNR(I,G) = 10\log_{10}(MAX_G^2 / MSE(I,G)) \quad (1)$$

$$MSE(I,G) = \frac{1}{mn}\sum_{i=1}^{m}\sum_{j=1}^{n}(I_{ij} - G_{ij})^2 \quad (2)$$

where $I$ and $G$ are respectively the estimated images and the ground truth images, the size of which is assumed to be $m*n$ pixels. *MSE* ($I$, $G$) represents the mean square error between $I$ and $G$. When the predicted image is closer to the actual images, the *MSE* is smaller so that *PSNR* becomes larger.

Therefore, the loss function (aim to minimize) related to *PSNR* is decided as $-1*PSNR + c$, where $c$ is a certain constant term. It should be mentioned that all the input, output, and supervisory data are normalized so that the maximum value ($MAX_G$) is always 1.

The error between the output permittivity $O$ and the true permittivity $T$ is assessed with $MSE$ ($O$, $T$). The smaller the value of the $MSE$ ($O$, $T$), the higher the accuracy of the DL model forecasting. Thus, the loss function about the permittivity estimation is directly demonstrated as $MSE$ ($O$, $T$). Meanwhile, for the classifier estimating the probabilities of the targets' quantity being, the cross-entropy (*CE*) function is employed as metrics. The *CE* of two probability distributions $P$ and $Q$ about the random variable $x$ are interpreted by:[39]

$$CE(P,Q) = -E_{x\sim P}[\log Q(x)] \tag{3}$$

where $P$ is the probability distribution of the labels (supervisory data), and $Q$ is the distribution of the model's prediction. $E_{x\sim P}$ is the expected value concerned with distribution $P$. In this work, since the distributions $P$ and $Q$ are discrete, the *CE* is simplified as:[40]

$$CE(P,Q) = -\sum_{x\in X} P(x)\log Q(x) \tag{4}$$

where $x$ is the set of all possible events. Supposing $Q$ manifests as [0.1 0.8 0.1], it means the probabilities of the target's quantity being 1, 2, and 3 are 0.1, 0.8, and 0.1, respectively. Under this condition, the target's quantity is most likely 2. Similarly, there exist three possible circumstances of labels $P$, which are [1 0 0], [0 1 0], and [0 0 1]. The three distributions stand for definite (100% probability) single, double, and triple targets, respectively. Actually, the *CE* loss function is continuously reduced to get an accurate prediction.

In summary, these three loss functions are eventually combined with distinctive weights ($\alpha$, $\beta$, and $\gamma$) as the total loss function of the entire model:

$$Loss_function = -\alpha * (PNSR(I,G) + c) + \beta * MSE(O,T) + \gamma * CE(P,Q) \tag{5}$$

The goal of the DL model training is to minimize this comprehensive loss function and achieve the desired accuracy of the three tasks. Consequently, the DL model can simultaneously perceive the quantity and the permittivity of the objects, as well as draw their super-resolution images. This multifunctional characteristics can provide an essential avenue to efficiently extract the target information.

## 3. RESULTS

In the following procedure, it is worth mentioning that all the model output results are based on the randomly packed training, validating, and test dataset. In order to ensure correctness, we randomly assign the dataset five times (conditions 1~5) and apply the assigned data for training, validation, and testing.

**3.1. Original DL model results**. To be specific, the DL model is saved and fixed for the ensuing performance test after training and validating with 700 epochs. Take dataset condition 1 as an instance, Figure 5a explicitly illustrates the *PSNR* curves after each training and validation epoch. As the epoch increases, it demonstrates that both the PSNR of the training dataset and validation dataset converge to a high value so that the model training is regarded as completed. Figure 5b illustrates the *MSE* between the predicted and true permittivity (*MSE*_permi), and the *MSE* between the predicted and true target quantities (*MSE*_peak). It is noted that all of the *MSE*s decrease to minimal values and then keep stable. Actually, according to the performance of all the training and verification curves, the DL model has produced satisfactory results after 700 epochs, indicating the model can be fixed and utilized for testing. Then, the model is fed with a test dataset containing 509 samples, while the predicted results of the test set are evaluated with several metrics in Table 1.

Firstly, the *PSNR* (*I*, *G*) and *MSE* are employed for discriminating of super-resolution images. As illustrated in Table 1, the *PSNR* of the test datasets in the five data divisions can reach at least 42dB, while the *MSE* is no more than $1e^{-4}$. These results illustrate the negligible differences between the genuine super-resolution images and the model-rebuilt images. More intuitively, as shown in Figure 5c-h, super-resolution images of five randomly-selected test samples in condition 1 are presented. The black curves are the model-predicted images and the cyan areas are the regions between the actual ground truths and the *x*-axis. The results verify that the images can be precisely reconstructed and the resolvable distance is attainable to approximately $0.2\lambda$, validating its powerful super-resolution imaging ability.

Regarding the detection of permittivity, the criteria are the *MSE* (*MSE*_permi) between the model's output permittivities and the truth, accompanied by the prediction accuracy (*Acc*_permi). In the five dataset division conditions, all the *MSE*_permis are lower than 0.07, and the prediction *Acc*_permis are 95.4% or higher. It should be noted that the calculation of the *Acc*_permi is to round the predicted permittivity value and then check if it is consistent with the actual permittivity.

Thus, the *Acc*_permi is established as a rough assessment of the data. To vividly demonstrate the distribution of anticipated permittivities, Figure 5i depicts the violin plot with the abscissa being the actual permittivities, while the ordinate represents the predicted permittivities. It is clarified that the majority of the anticipated data is highly concentrated, which is very close to the corresponding actual value, particularly when the permittivity belongs to 10~17 or 20~21. However, the distribution of the predicted permittivity is relatively divergent when the actual permittivity is 18 or 19. A few predictions have an error of approximately 10%. Nevertheless, large deviations occur only in isolated cases, and the overall permittivity predictions are relatively accurate.

As for the sensing of the target quantity, the accuracies of the test datasets reach 100% and the *MSE*_peak is almost zero for all the five data division conditions in Table 1. These results demonstrate that the estimation of the targets' quantity reaches high precision.

To sum up, it can be seen that the multi-task DL model proposed in this work can simultaneously perceive the object information and plot super-resolution images with relatively high precision. Therefore, with the addition of a DL network, target information reconstruction under multi-harmonic illumination is achieved. Due to the data-driven underlying logic, the DL model is applied in this complex electromagnetic environment that is complicated for conventional analytical methods or iterative algorithms. To some extent, this methodology also provides an alternative way for other electromagnetic-related problems.

**Table 1.** Test results of the DL model with the fit dataset

| Condition | PSNR(dB) | MSE_image | MSE_permi | MSE_peak | Acc_peak | Acc_permi |
|---|---|---|---|---|---|---|
| 1 | 42.0793 | 1e$^{-4}$ | 0.0641 | 2.342 e$^{-10}$ | 100% | 95.68% |
| 2 | 44.0001 | 5.5203 e$^{-5}$ | 0.0668 | 0 | 100% | 95.48% |
| 3 | 43.5245 | 7.6389 e$^{-5}$ | 0.0661 | 0 | 100% | 96.86% |
| 4 | 43.2255 | 8.9319 e$^{-5}$ | 0.0583 | 7.0261 e$^{-10}$ | 100% | 95.68% |
| 5 | 42.7303 | 7.4370 e$^{-5}$ | 0.0391 | 1.3852e$^{-5}$ | 100% | 96.46% |

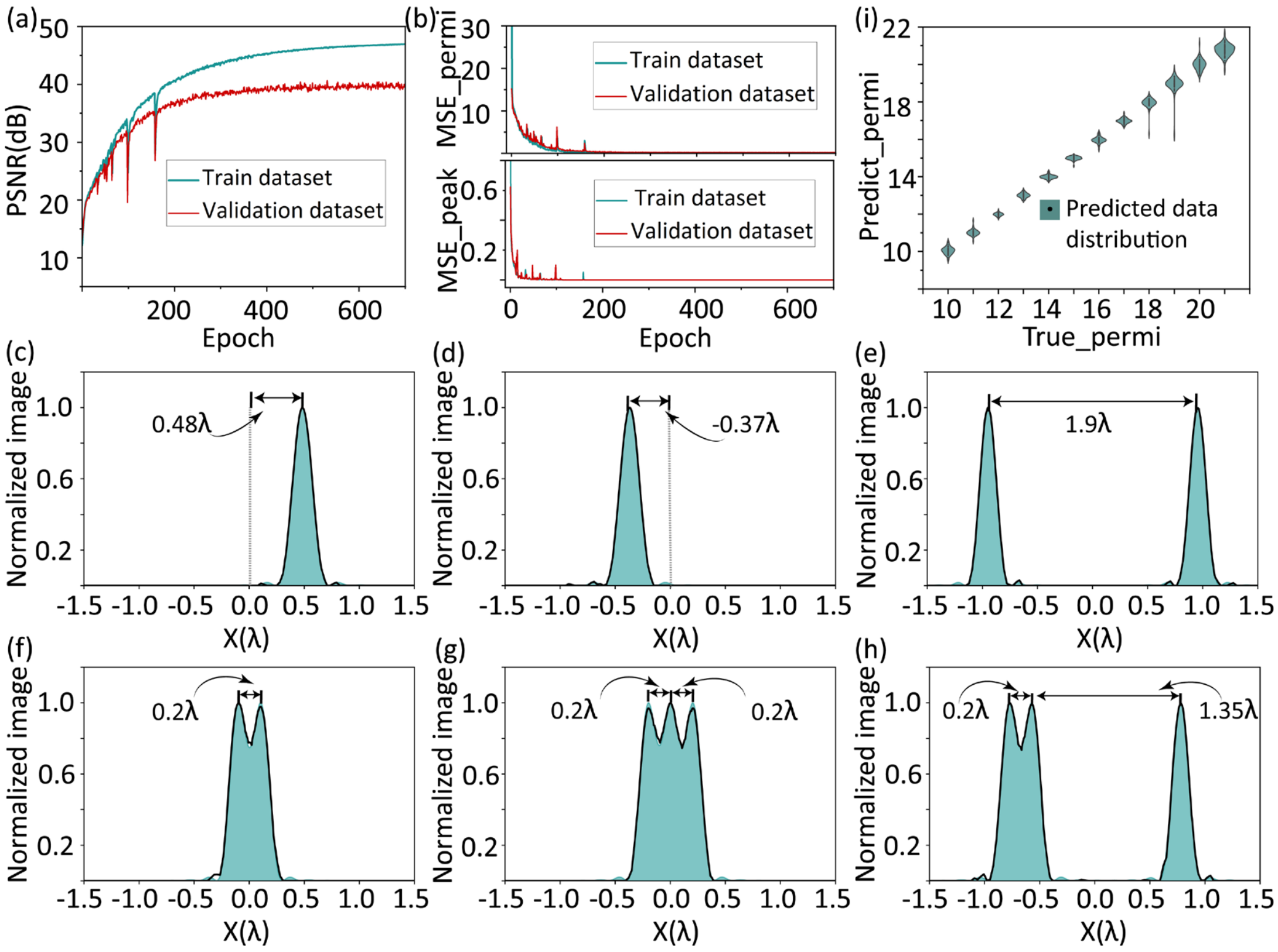

**Figure 5.** Results of the proposed multitask deep learning (DL) model on the original dataset. (a) *PSNR* of the training and validating dataset when the epoch increases. (b) Training and validating *MSE* of the permittivity (upper panel) and quantity (lower panel) prediction. (c-h) Super-resolution images of five randomly-selected test samples, where the black solid lines represent the model-predicted images, and the areas between the actual image curves and the *x*-axis are cyan. (i) The distribution of the model-output permittivity values related to the corresponding true permittivity values.

**3.2. Results without fitting input data.** In the aforementioned verification, the DL network's input data is a smooth curve, which is fitted by the detected 21 discrete points. Actually, the acquired initial data is only the 21 discrete points, and the fitting method is a post-processing strategy. Thus, it is crucial to figure out whether the main information learned by the DL network is from the discrete points or the subsequent smooth curve fitting procedure. For this purpose, a direct interpolation on these 21 discrete points is employed to obtain a non-smooth curve. The obtained line curves are fed into the DL network, afterward realizing the training, validating, and testing of the DL model. Similarly, five different data divisions are exerted to exclude the accidental results. The sequence of the randomly shuffling data in conditions 1~5 are respectively identical (for example, condition 1 in Tables 1 and 2 are the same). Consequently, the influence of different dataset divisions can be removed when the fitting and interpolation post-processes are compared.

Take condition 1 as an example, the learning curves of the three tasks are illustrated to be converged after 700 epochs, as plotted in Figure 6a,b. The test results of the DL model are given in Table 2. Comparing the condition 1 results between Tables 1 and 2, the *PSNR* (*I*, *G*) in Table 2 is slightly decreased with a 39.79dB value, while the *MSE*_image gets doubled. Figure 6c-h presents the same test samples as those in Figure 5c-h, where the black line is the predicted image and the area between the true image and the *x*-axis is occupied with cyan. It is obvious that there are some differences between the estimated images and the ground truths in Figures 6c,f,g, especially in Figure 6f. This phenomenon indicates that the super-resolution image prediction based on the interpolation data has an increased error. For the permittivity estimation, Figure 6i presents that the forecast is concentrated on the truth value, although the *MSE*_permi and *Acc*_permi are marginally worse in Table 2. As for the targets' quantity prediction, the *Acc*_peaks are always 100%, which demonstrates that this task is hardly influenced by the data post-process method. The DL model's performance is similar for the five data division conditions.

To sum up, the effect of the DL model is also available when the input data is interpolated with discrete points. However, in contrast to the fit-data-supported DL model, the interpolated-data-enabled model performs inadequately on super-resolution image generation. Therefore, to some extent, the model-learned information is mainly from the 21 discrete points, and a limited amount of information is related to the post-processing method.

**3.3. Results with reducing input information.** Aside from the influence of the discrete data post-process method, the field detected range may be significant for the network performance. When the detection range is decreased, the length of the input curve gets shorter, meaning fewer learnable discrete points. Resulting of the former section, the main information that the DL network learns is contained in the 21 discrete points. The reduction of the learnable discrete points manifests the decreased information provided to the DL network, which may lead to a weak behavior of the network. To confirm this suggestion, the input data (intensity field distribution) length is set as $7\lambda$, $5\lambda$, and $3\lambda$. Table 3 illustrates the behaviors of the interrelated networks' performances on the test datasets (the dataset is randomly assigned the same as condition 1 in Tables 1 and 2).

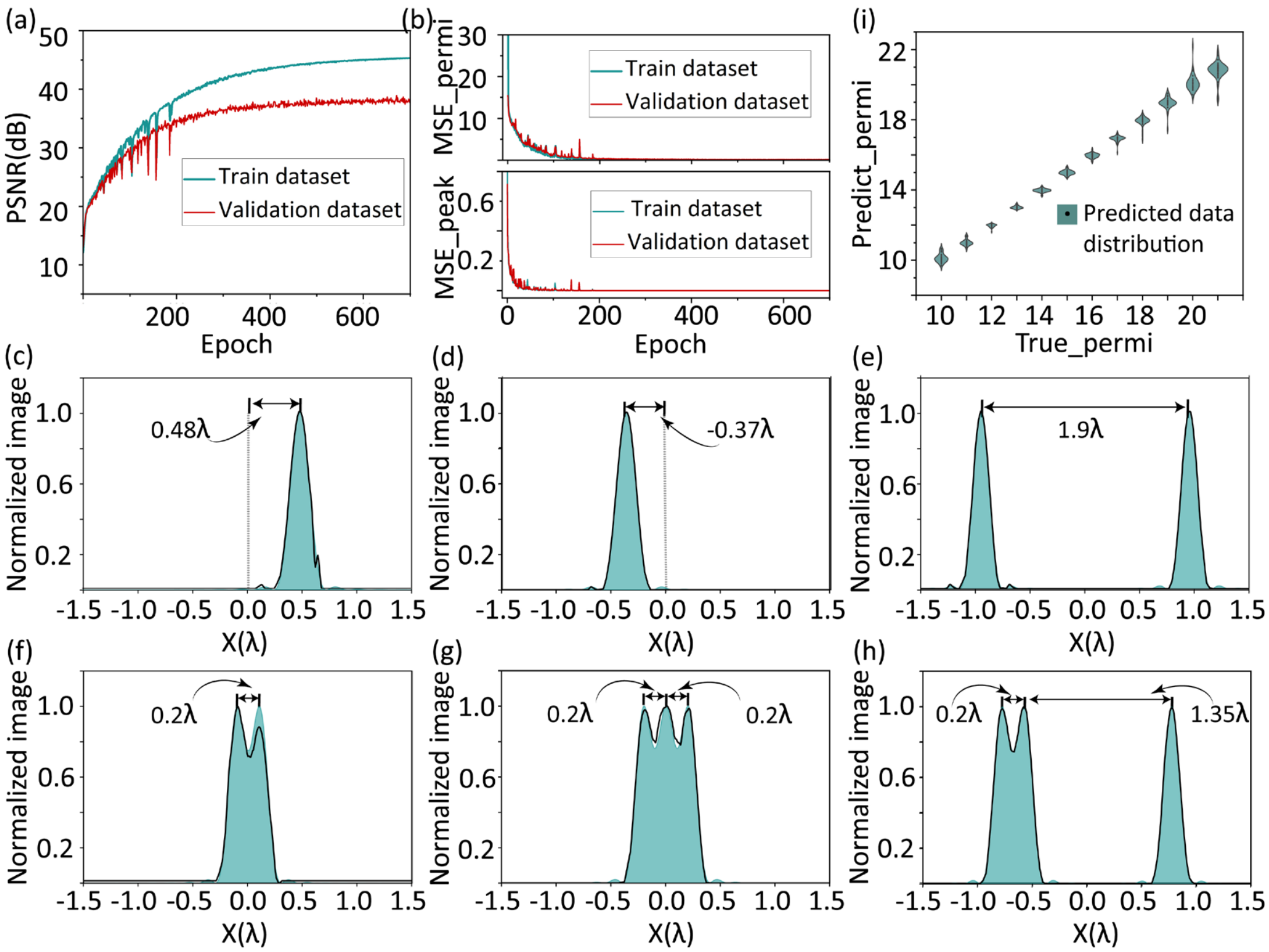

**Figure 6.** Results of the updated multitask DL model on the interpolated dataset. (a) *PSNR* with increasing epoch. (b) *MSE* between the prediction and the truth, where permittivity is in the upper panel and quantity is in the lower panel. (c-h) The model output images (black solid lines) and the areas between actual super-resolution image curves and the *x*-axis (cyan areas). (i) The model-estimated permittivity values and their corresponding true values.

**Table 2.** Test results of the DL model with the interpolated dataset

| Condition | PSNR(dB) | MSE_image | MSE_permi | MSE_peak | Acc_peak | Acc_permi |
|---|---|---|---|---|---|---|
| 1 | 39.7889 | $2e^{-4}$ | 0.0739 | 0 | 100% | 94.30% |
| 2 | 43.3246 | $7.5167\ e^{-5}$ | 0.0922 | 0 | 100% | 95.09% |
| 3 | 43.1969 | $9.0135e^{-5}$ | 0.1020 | 0 | 100% | 96.27% |
| 4 | 41.1968 | $1e^{-4}$ | 0.0816 | $2.8104e^{-9}$ | 100% | 95.28% |
| 5 | 42.0748 | $9.3789e^{-5}$ | 0.0734 | $4e^{-4}$ | 100% | 95.48% |

As shown in Figure 7a,b, along with the decline of the input data length, the *PSNR* (*I*, *G*) becomes smaller and the *MSE*_image gets larger. In this manner, the image reconstruction ability of the DL model becomes worse and the accuracy is lower. When the length is 3$\lambda$, the *PSNR* (*I*, *G*) is only 32dB. Analogously, the accuracy of the permittivity estimation also shrinks as the input length drops. Figure 7b,c presents the trend of the *MSE*_permi and *Acc*_permi, which shows that the networks' performance declines dramatically when the data length reduces from 5$\lambda$ to 3$\lambda$. For the targets' quantity prediction, the accuracy is almost stable with a tiny variety, demonstrating this task is relatively easy and can be realized efficiently with brief information.

Overall, performances of the network get worse when the input intensity curves are shorter, due to the reduced effective information. In another aspect, a large input of information not only causes admirable network performance but also increases the cost of data acquirement. Thus, the 10$\lambda$ detection range containing 21 discrete points may be a reasonable value that can guarantee impressive DL model performance, without requiring large data acquisition costs.

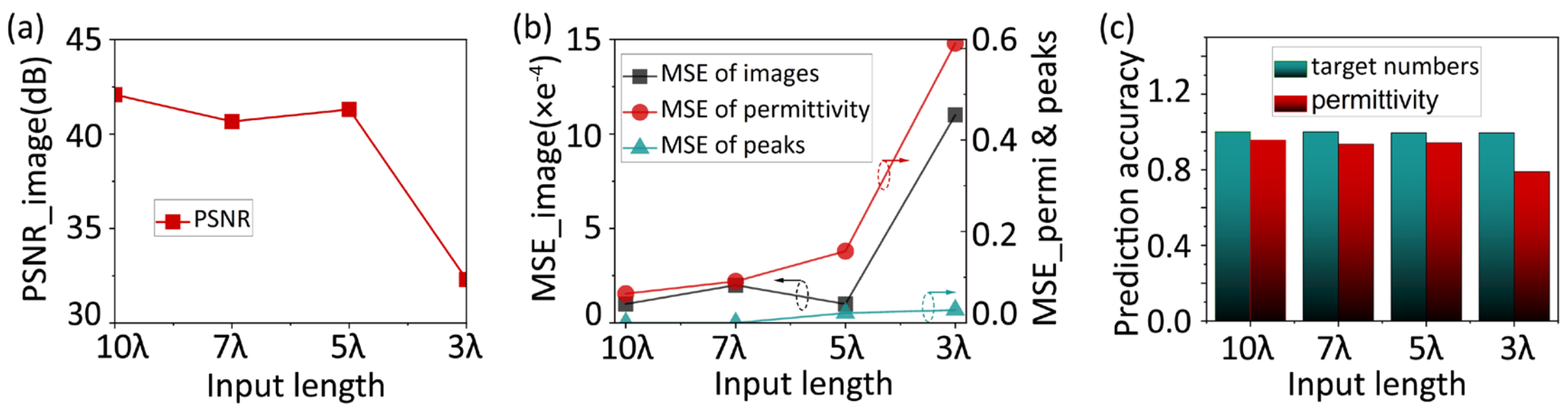

**Figure 7.** Results of the DL models with the different input information. (a) *PSNR* between the predicted images and the actual super-resolution images. (b) *MSE* between the predictions (targets' images, permittivity values, and quantities) and truths. (c) Prediction accuracies of the targets' quantities and permittivity values.

**Table 3.** Test results of the DL model with reduced information

| Sampling | PSNR(dB) | MSE_image | MSE_permi | MSE_peak | Acc_peak | Acc_permi |
|---|---|---|---|---|---|---|
| 10λ | 42.0793 | $1e^{-4}$ | 0.0641 | $2.342\ e^{-10}$ | 100% | 95.68% |
| 7λ | 40.6604 | $2e^{-4}$ | 0.0908 | $1.2302e^{-6}$ | 100% | 93.52% |
| 5λ | 41.3006 | $1e^{-4}$ | 0.1571 | 0.0218 | 99.61% | 94.30% |
| 3λ | 32.3082 | $1.1e^{-3}$ | 0.6115 | 0.0282 | 99.61% | 78.98% |

**3.4. Results with the different loss functions.** In the all aforementioned DL networks, the total loss function is demonstrated in equation (5), which is composed of $1 - PSNR\ (I, G)$, *MSE* (*O*,

*T*), and *CE* (*P*, *Q*) in different weights. *PSNR* (*I*, *G*) is employed as a description for the consistency between the reconstructed image *I* and the actual super-resolution image *G*. However, another generally applied metric is *MSE* (*I*, *G*), which can also exhibit the distance between the *I* and *G*. To verify the effect of *MSE* (*I*, *G*), the 1 − *PSNR* (*I*, *G*) item is instead by *MSE* (*I*, *G*) to train another new DL model. In this condition, the updated total loss function is expressed as follows:

$$Loss_function' = \alpha' * MSE(I,G) + \beta' * MSE(O,T) + \gamma' * CE(P,Q) \tag{6}$$

Utilizing the revised loss function, the model's performance in the test dataset is shown in Table 4 (dataset division follows the previous condition 1). Both the *PSNR* and *MSE* are the criteria that demonstrate the models' effects. In Table 4, after applying the *MSE* (*I*, *G*) as an item of the loss function, the *PSNR* decreases and the *MSE*_image becomes three times the original. This phenomenon reveals that the DL model performs worse in super-resolution image reconstruction. The accuracy rate of permittivity prediction is still relatively high, which is almost not affected by the loss function. Similarly, the forecast of the target quantity is also not sensitive to the change of the loss function, which constantly has a 100% accuracy ratio.

According to the foregoing analysis, applying the *MSE* (*I*, *G*) as part of the loss function is relatively deficient for the DL model in this work. In contrast, the DL model trained by the original loss function with 1 − *PSNR* (*I*, *G*) is more precise, especially in the super-resolution image recovery task.

**Table 4.** Test results of the DL model with different loss functions

| Item In loss | PSNR(dB) | MSE_image | MSE_permi | MSE_peak | Acc_peak | Acc_permi |
|---|---|---|---|---|---|---|
| 1 – PSNR | 42.0793 | $1e^{-4}$ | 0.0641 | $2.342\ e^{-10}$ | 100% | 95.68% |
| MSE | 38.3349 | $3e^{-4}$ | 0.0960 | $4.6841e^{-10}$ | 100% | 94.89% |

## 4. DISCUSSION

This work employs a DL network to conduct the target's feature perceptions and information retrievals based on a single intensity curve. We can reliably acquire the target's permittivity, quantity, and super-resolution images due to the network's effective and accurate post-processing capabilities. Instead of continually updating the solution to reach convergence in the iterative method, the DL network reconstructs the information instantly once the training procedure is complete. Hence, to some extent, DL networks offer great potential and superiority in terms of fast and precise post-processing.

Despite the focus of this study is employing DL models to retrieve target information from their scattered fields, the DL methodology may also be applied to directly forecast the scattered fields of the already-known targets. In the following, a new DL model (DL model 2) is proposed to predict the scattered field distribution in the multi-harmonic illumination. With the assistance of DL model 2, part of the scattered intensity distribution can be obtained without simulation software, thus considerably reducing the time required for data collection. In this condition, the intensity field distribution predicted by DL model 2 can be utilized as training data for the aforementioned target reconstruction problem (DL model 1). This confirmation highlights the advantages and benefits that introducing DL models in electromagnetic scenarios.

Figure 8 illustrates the schematic diagram of the DL model 2. The illumination is still the multi-harmonic mode supported by the circular grating metasurface. In the simulation, multiple harmonics waves excited by incident plane waves are irradiated on the targets. Then, the intensity distribution of the scattered field is obtained on the detection line. Actually, this process can be predicted by the elaborately-designed DL model 2. As demonstrated in Figure 8, the quantity, permittivity, and super-resolution image of targets are fed into the model, and then processed by the multi-layer network. More specifically, the input super-resolution images first pass through a multi-layer CNN network, while the input target quantity and permittivity are dealt by dimension-enhancement networks. Next, the outputs of the above three modules are pieced together into a tensor and dimensionally reduced with the FC layers. Then, the modified tensor undergoes a trans-convolution network and is output as the final predicted scattering intensity field. A total of 3062 sets of data are randomly chosen to train the DL model 2, 1019 sets of data are selected as the validation, and the remaining 1019 datasets are utilized as the test set (the ratio of training, validation, and test dataset is 3:1:1). Finally, the average PSNR between the model-predicted results and the actual scattered fields is up to 50dB, which verifies the precise and efficient scattered field prediction ability of the model 2.

In order to graphically display the effect of the model, six test samples are randomly selected and demonstrated in Figure 9. Red dotted lines in these diagrams indicate the super-resolution images of the target, *P* refers to the target's permittivity, *N* is the number of targets, and the black solid lines represent the model's envisaged intensity field distribution. Additionally, the area between the actual scattered field distribution curve and the $x$-axis is filled with cyan shade. The predicted field of binary targets with a permittivity of 20 and a $0.32\lambda$ separation is plotted in Figure 9b. The cyan area's envelope and the expected intensity field are practically identical. Similarly, Figure 9c-f illustrates the model-anticipated and actual field distributions of multi-target with various positions and permittivities. It is worth noticing that there are just some minor variances in a few regions and the deviation between the two field distributions is tiny.

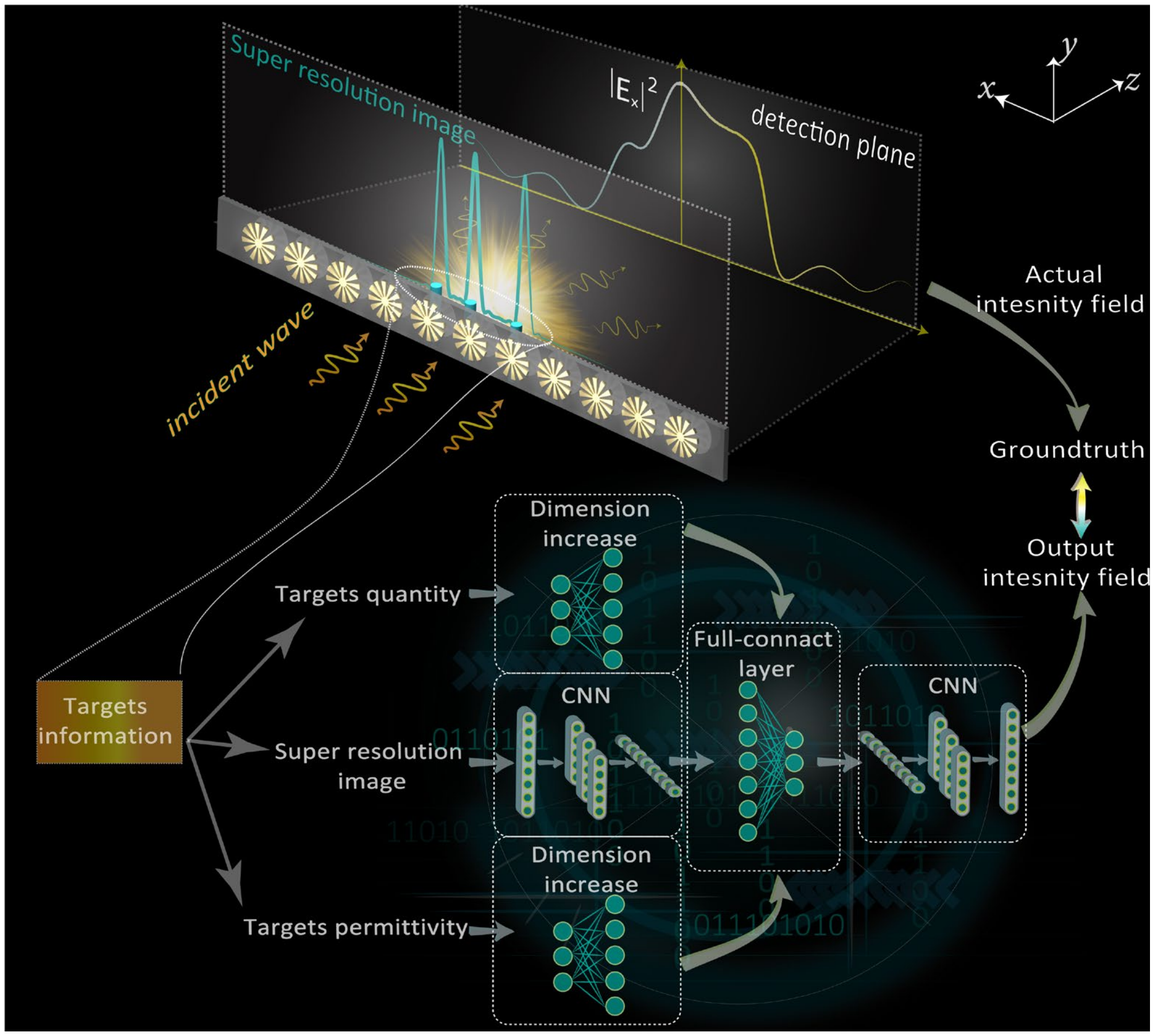

**Figure 8.** Schematic of the scattering electric field prediction with the assistance of a DL model.

In conclusion, the DL network model 2 is capable of estimating the scattered intensity field of the already-known targets, while achieving excellent accuracy rates. The execution of this strategy demonstrates that the DL method can not only be applied to solve the reconstruction problem (perceiving the target's characteristics and super-resolution images) but also can evaluate the diffraction and scattering of electromagnetic waves in diverse illumination environments. On the other hand, without the DL model 2, simulation software is generally utilized to obtain all of the datasets to support the training and test of DL model 1. To obtain all the 5100 sets of data, approximately 430 hours are requested. While introducing the DL model 2, 1019 sets of data can be directly generated through the DL model 2. Therefore, the adoption of DL model 2 may greatly improve the efficiency of data generation.

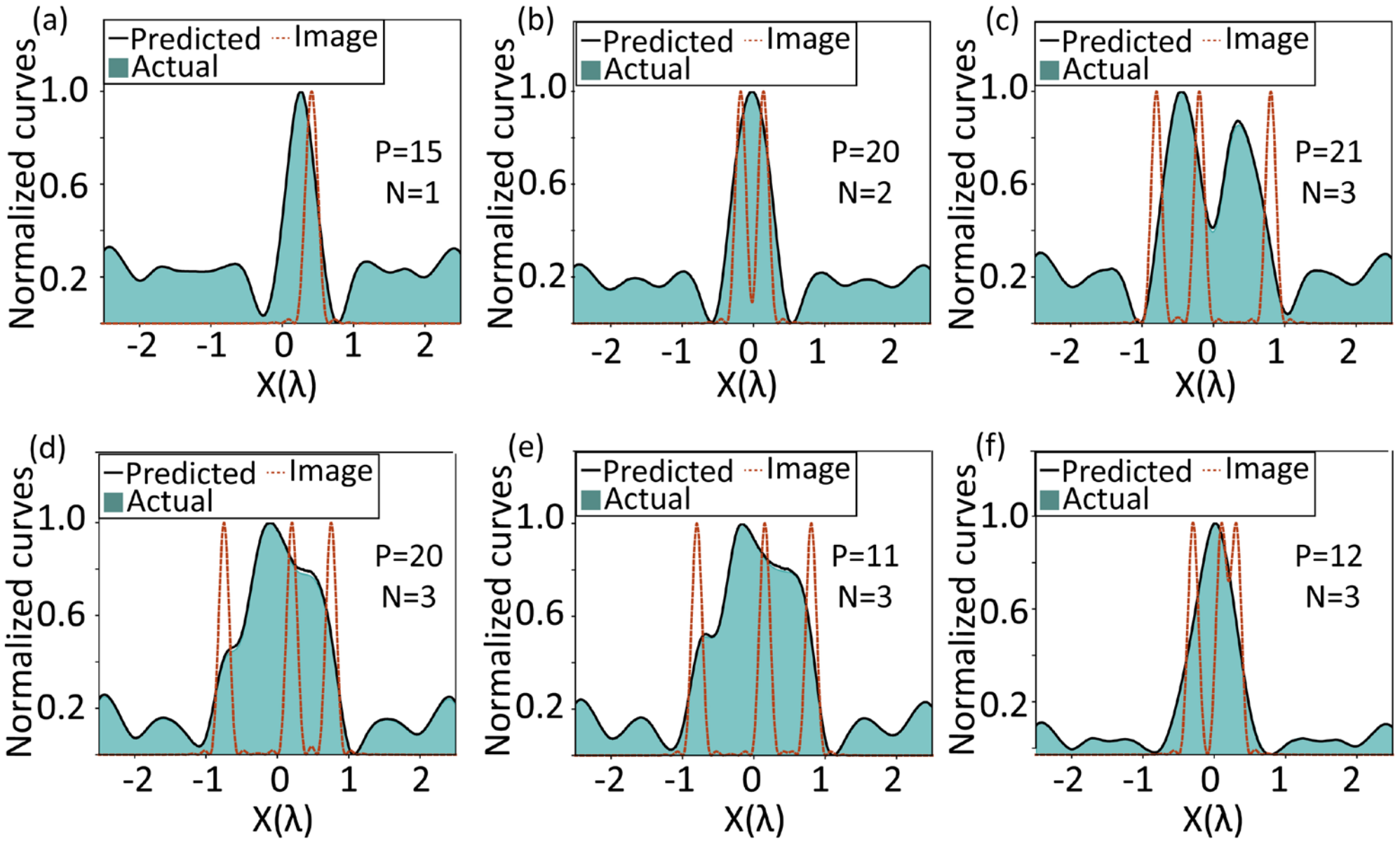

**Figure 9.** Results of the DL model 2 for the prediction of electric field distributions. (a-f) Different randomly-chose test results, presenting the predicted field distributions (black solid lines) and the actual field distributions (areas between the actual field distributions and the *x*-axis are cyan). The red dotted lines are the corresponding super-resolution images of the targets. *P* is the targets' permittivity values and *N* represents the targets' quantity.

The introduction of the DL network allows for not only the inverse reconstruction of unknown targets but also the forward prediction of the scattered field. Thus, it can be applied to systems that are laborious or time-consuming to solve by conventional analytical methods or algorithms, while also attaining high accuracy to some extent.

## 5. CONCLUSION

In this work, a DL network model is proposed to simultaneously sense and reconstruct the unknown targets under multi-harmonic illumination. The multi-harmonic illumination is realized with the periodic metasurface, and the effectiveness of the target introduction is verified in both simulation and experimentation. The DL network is composed of the encoder, decoder, classifier, and regressor. Feeding the intense electric field in the network, it can predict the super-resolution images, permittivity, and quantity of the targets with relatively high precision and accuracy (*PSNR* > 42dB, *Acc*_peak = 100%, and *Acc*_permi > 95%). Compared with the analytical solution in SIM[41], the deep learning method can obtain super-resolution images based on only one frame of electric field intensity. In contrast to the SIM employing the iterative GS algorithm which is restricted by a single-harmonic illumination[24, 37], this work can be effective in multi-harmonic illumination circumstances. Moreover, apart from the super-resolution image, this approach simultaneously senses the quantity and permittivity of the targets, achieving multifunctional reconstructions. In brief, the attachment of this DL model provides a promising way to precisely and effectively reconstruct targets' information.

To further save the data acquirement time, another new DL model is established to produce the electric field distribution with the targets' information being the model's inputs. In this condition, 20% of the dataset can be generated by the DL model, without the software simulation. To sum up, the proposed two DL models demonstrate that a deep learning approach is a powerful tool for a range of electromagnetic inverse reconstruction or forward prediction challenges.

## AUTHOR INFORMATION

### Corresponding Author

**Pu-Kun Liu** − *State Key Laboratory of Advanced Optical Communication Systems and Networks, School of Electronics, Peking University, Beijing 100871, China;* ORCID: 0000-0001-8750-7899; Email: pkliu@pku.edu.cn.

### Author Contributions

The manuscript was written through the contributions of all authors. All authors have given approval to the final version of the manuscript.

### Notes

The authors declare no competing financial interest.

## ACKNOWLEDGMENT

The authors would like to thank Fan-Hong Li, Peking University for the help during the experiment process. Thanks for the valuable advice to analyze the model from Dr. Tie-Jun Huang and Dr. Li-Zheng Yin. This work was supported by the National Natural Science Foundation of China under Grant No. 61971013 and the National Key Research and Development Program of China under Grant No. 2019YFA0210203.